\documentclass[aps,pra,twocolumn,groupedaddress,showpacs]{revtex4}

\usepackage{graphicx}
\usepackage{dcolumn}
\usepackage{bm}
\usepackage{verbatim}
\usepackage{amsmath,amssymb}
\usepackage{relsize,exscale}
\usepackage{color}
\usepackage{hyperref}
\begin{document}

\title{Cloning in nonlinear Hamiltonian quantum and hybrid mechanics}

\author{D. Arsenovi\'c}
\author{N. Buri\'c}
\email[]{buric@ipb.ac.rs}
\author{D. B. Popovi\'c}
\author{M. Radonji\'c}
\author{S. Prvanovi\'c}

\affiliation{Institute of Physics, University of Belgrade, Pregrevica 118, 11080 Belgrade, Serbia}

\begin{abstract}
Possibility of state cloning is analyzed in two types of generalizations of 
quantum mechanics with nonlinear evolution. It is first shown that nonlinear 
Hamiltonian quantum mechanics does not admit cloning without the cloning 
machine. It is then demonstrated that the addition of the cloning machine, 
treated as a quantum or as a classical system, makes the cloning possible by 
nonlinear Hamiltonian evolution. However, a special type of quantum-classical theory,
known as the mean-field Hamiltonian hybrid mechanics, does not admit cloning by
natural evolution. The latter represents an example of a theory where it appears to
be possible to communicate between two quantum systems at super-luminal speed, but
at the same time it is impossible to clone quantum pure states.
\end{abstract}


\maketitle

\section{Introduction}

Impossibility of cloning unknown quantum states is a fundamental property of 
quantum systems \cite{Wooters,Gisin}. It has been used as a basis for information
theoretic axiomatization of quantum mechanics (QM) \cite{Bub},  and is crucial
in several quantum information processing tasks \cite{tasks}. 
Roughly speaking, state cloning is a process which involves at least two 
systems: an object system whose state is to be cloned and a target system whose 
state is transformed into the state which is equal to the state of the object 
system. Often, and in order to allow for the most general type of processes, one 
includes also an ancilla system, which is in the context of cloning called the 
cloning machine. Standard simple proofs of no-cloning involve properties of 
quantum processes, such as: a) linearity or b) preservation of a nontrivial 
distance between quantum states, and also use c) direct product structure of 
composite quantum systems. The properties a), b) and c) are not independent in 
QM, but each of them implies crucial differences between QM and classical
mechanics (CM). Modifying any of the three properties leads to generalization
of QM, which is also different from CM. Some of generalizations are mathematically
inconsistent or in conflict with other fundamental physical theories like special
relativity or thermodynamics \cite{Gisin1, Peres}. Depending on the modification,
cloning of states in the modified theory might, but need not, be possible.
Possibility of cloning in a modified theory need not be related with super-luminal
signaling, like it is in the standard QM. It is the purpose of this communication
to discuss possibility of cloning in two types of modifications of QM. Both types
of the modified theories are formulated in the framework of Hamiltonian dynamical
systems (HDS). Standard QM can be formulated as a linear HDS on an appropriate phase
space \cite{Ashteckar,Brody}. Mathematically consistent generalizations of QM can be
obtained by modifying some of the standard QM properties but remaining in the
framework of HDS. It is known that cloning is possible in classical mechanics with
Hamiltonian dynamics \cite{classical}. So, it is interesting to investigate the
possibility of cloning within different Hamiltonian generalizations of QM. The first
class of modified theories that we shall study retains all the kinematical properties
of QM in HDS formulation, but allows evolution given by general nonlinear
Hamiltonian equations. Weinberg \cite{Wein} and Bialynicki-Birula and Mycielski
\cite{Birula} nonlinear Schr\"{o}dinger equations are actually of this type.
We abbreviate this type of theories as NHQM standing for nonlinear Hamiltonian QM.
The second type of modified theories assumes that some of the degrees of 
freedom (DF) of the HDS corresponding to a bipartite system are constrained to 
behave as classical DF \cite{usPRA,Elze}. We call this type Hamiltonian 
hybrid mechanics (HHM). The constraint implies nonlinear evolution of both 
classical-like DF (CDF) and of quantum DF (QDF) \cite{usPRA,JaAnnPhys}, but also 
changes the way in which the phase spaces of QDF and CDF are composed to form 
the phase space of the total hybrid system. Thus, in this type of theories the 
evolution is nonlinear and the tensor product rule is not valid for all DF. Our 
main results are: a) self-replication, i.e., a type of cloning in the restricted
sense without the cloning machine, is impossible in NHQM; b) inclusion of a
{\it quantum} cloning machine makes the cloning in NHQM possible, and c) cloning
with the object and the target quantum systems and a {\it classical} cloning machine
is also possible with nonlinear hybrid evolution. Thus, these two types of nonlinear 
generalizations of quantum mechanics, in which the evolution of the total system 
is Hamiltonian, allow the cloning of quantum states by natural evolution. However,
cloning is impossible in a type of HHM with the Hamiltonian of a special mean-field
form. These results are to be contrasted with the known result that the cloning is
impossible within bipartite classical Hamiltonian systems (object and target), but
becomes possible within three-partite systems \cite{classical} (object, target and
cloning machine). In the latter case the cloning can be achieved by a linear symplectic
map \cite{classical}. Thus, it seems that if the object and the target are quantum
(tensor product) and the evolution of the total system that includes the machine is
Hamiltonian, then the cloning map is necessarily nonlinear, irrespective of the
quantum or classical nature of the cloning machine. However, if all the three systems
are classical (Cartesian product), then the cloning is possible by linear
transformations which are symplectic on the total phase space.

Structure of the paper is as follows. The next section serves to recapitulate,
very briefly, the Hamiltonian formulation of QM and of the HHM, and then to formulate
the definitions of the the cloning and self-replication processes in NHQM and in HHM.
In Section III we prove our main results concerning the cloning (and self-replication)
in NHQM and in HHM. Section IV contains several remarks which provide a discussion of
our results. Summary is given in Section V.

\section{Formulation of cloning in Hamiltonian quantum and hybrid theories}

\subsection{Hamiltonian formulation of QM and of hybrid mechanics}

{\it Hamiltonian QM and nonlinear generalizations}

Quantum and classical mechanics can be formulated using the same mathematical framework
of Hamiltonian dynamical systems $({\cal M},\omega,H)$ where ${\cal M}$ is a symplectic
manifold, $\omega$ the corresponding symplectic structure and $H$ the Hamilton's function.
Formulation of the classical mechanics of isolated conservative systems using 
$({\cal M},\omega,H)$ is standard \cite{Arnold}. The formulation of quantum mechanics
in terms of $({\cal M},\omega,g,H)$, where $g$ is an appropriate Riemann structure, is
perhaps less well known, but shall not be presented here in any detail since there exist 
excellent reviews \cite{Ashteckar,Brody}. Very briefly, the basic observation beyond the 
Hamiltonian formulation of quantum mechanics is that the evolution of a pure quantum
state in a Hilbert space ${\cal H}^{N}$, given by the Schr\"odinger equation, can be
equivalently described by a HDS on an Euclidean manifold ${\cal M}={\mathbb R}^{2N}$.
Here $N$ is the complex dimension of the relevant Hilbert space. The manifold ${\cal M}$
is just the Hilbert space considered as a real manifold, with the symplectic and Riemann
structures given by the real and the imaginary parts of the Hilbert scalar product. The
manifold also possesses an almost complex structure $J^2=-I$ such that $g(x,y)=\omega(x,Jy)$.
Normalization and global phase invariance of quantum states can be incorporated into
the formulation of the phase space of quantum states which is the projective space 
$P{\cal H}^{N-1}\sim S^{2N-1}/S^1$, with the corresponding symplectic, Riemann and
almost complex structures. However, in our computation we shall use the Hamiltonian 
formulation based on ${\mathbb R}^{2N}$, so that when treating the problem of cloning,
we shall have to take care of the global phase invariance explicitly. Representing
a normalized vector $|\psi\rangle\in {\cal H}^{N}$ in an arbitrary basis
$\{|e_j\rangle\}_{j=1}^{N}$ as $|\psi\rangle=\sum_{j=1}^{N} c_j|e_j\rangle$,
one can introduce the real canonical coordinates $x_j=(\bar{c}_j+c_j)/\sqrt{2\hbar}$,
$y_j=i(\bar{c}_j-c_j)/\sqrt {2\hbar}$, $j=1,2,\dots,N$, where bar indicates complex conjugation.
Change of the basis by a unitary map involves a linear symplectic transformation of the canonical
coordinates. Generic point from ${\cal M}$ will also be denoted by $X$ or $X^a$, where $a=1,2,\dots,2N$
is an abstract index, such that $X^a=x_a$, $a=1,2,\dots,N$ and $X^a=y_a$, $a=N+1,\dots,2N$.
If we want to stress that the point $X$ corresponds to the vector $|\psi\rangle\in{\cal H}^{N}$
we will write $X_{\psi}$, and vice versa $|\psi_{X}\rangle$ for the vector corresponding to 
the point $X$. It should be stressed, perhaps, that the canonical coordinates $(x_j,y_j)$ have
nothing to do with the canonical coordinates of the classical system that after quantization
gives the considered quantum system with the Hilbert space ${\cal H}^{N}$. The Hamilton's
function $H(X)$ is given by the quantum expectation of the Hamiltonian $\hat H$ in the
state $|\psi_X\rangle$: ${H(X)}=\langle\psi_X|\hat H|\psi_X\rangle$. 
The Schr\"odinger dynamical law is that of Hamiltonian mechanics
\begin{equation}\label{Scheq}
\dot X^a=\omega^{ab}\nabla_b H.
\end{equation}
where $\omega^{ab}$ is the standard unit symplectic matrix
\begin{equation}\label{symp}
\omega=\begin{pmatrix} \mathbf{0}&\mathbf{1}\cr
-\mathbf{1}&\mathbf{0} \end{pmatrix},
\end{equation}
where $\mathbf{0}$ and $\mathbf{1}$ are zero and unit matrices of dimension $N$.

In the Hilbert space QM and in the Hamiltonian CM the dynamical variables can be
introduced formally as generators of the isomorphisms of the respective relevant 
structures. In QM these are self-adjoined operators generating unitary transformations
that preserve the Hilbert scalar product. In the Hamiltonian formulation of QM the
Hilbert scalar product generates both the symplectic and the metric Riemann structures.
The symplectic structure is preserved by Hamiltonian vector fields of arbitrary smooth 
functions, but the metric is preserved only by the Killing vector fields, i.e., by the
Hamiltonian vector fields generated by quadratic functions of the canonical variables.
In particular, the unitarity of the QM evolution implies that the Hamilton equations
(\ref{Scheq}) are linear. All observables are represented by quadratic functions $A(X)$
on ${\cal M}$ and are the quantum mechanical expectations of the corresponding quantum
observables $A(X)=\langle\psi_X|\hat A|\psi_X\rangle$. On the other hand, the canonical
coordinates of the quantum phase space do not have physical interpretation. It is
important to observe that the Poisson bracket between two quadratic functions is also a 
quadratic function and satisfies
\begin{equation}\label{Poisson}
\{A_1(X),A_2(X)\}=\frac{1}{i\hbar}\langle\psi_X|[\hat A_1,\hat A_2]|\psi_X\rangle.
\end{equation}

In what follows we shall need to consider a bipartite quantum system composed of two
systems with Hilbert spaces ${\cal H}_1^{N_1}$ and ${\cal H}_2^{N_2}$. The phase space
of the total system is the manifold ${\cal M}_{12}={\mathbb R}^{2{N_1 N_2}}\sim
{\cal H}_1^{N_1}\otimes {\cal H}_2^{N_2}$. Of course, the space ${\cal M}_{12}$ is
much larger than the Cartesian product ${\cal M}_1\times {\cal M}_2$, which is relevant
for the formation of classical compound systems. If $|e^1_j\rangle$ and $|e^2_k\rangle$
are basis vectors in ${\cal H}_1^{N_1}$ and ${\cal H}_2^{N_2}$ respectively, with the
corresponding canonical coordinates $(x^1_j,y^1_j)$ and $(x^2_k,y^2_k)$, then the
canonical coordinates $(x^{12}_l,y^{12}_l)$ corresponding to the basis $|e_j^1\rangle
\otimes|e_k^2\rangle$ in ${\cal H}_1^{N_1}\otimes {\cal H}_2^{N_2}$ are given by rather
complicated formulas in general. Fortunately, we shall need only the formulas in the
most simple cases, further simplified by a special choice of the target system state
before the cloning transformation. In what follows we shall denote the composition of
phase spaces of two systems with phase spaces ${\cal M}_1$ and ${\cal M}_2$ by
${\cal M}_1\odot{\cal M}_2$, which means ${\cal M}_{12}$ in the quantum and 
${\cal M}_1\times {\cal M}_2$ in the classical case.

The Hamiltonian formulation of QM suggests natural formal generalizations \cite{Ashteckar}.
Several such generalizations could be seen as special cases of the theory called extended
quantum mechanics which was introduced and extensively studied in \cite {Bona}. The most
obvious one is to consider a theory where the evolution can be generated by functions 
which are not quadratic \cite{Ashteckar,Mielnik,Wein}, but to retain the assumption that
only the quadratic functions correspond to physical observables, and to retain the 
composition rule for compound systems. This would correspond to a nonlinear Schr\"odinger
evolution equation. Such a theory, which we abbreviate by NHQM, is still a HDS with the
same set of states and observables as in QM, but the Hamiltonian evolution equations are 
nonlinear and the metric is not evolution invariant. Since the proofs of no-cloning property
in QM are based on linearity or unitarity of the QM evolution, it is interesting to
investigate if the cloning is possible in NHQM.

{\it Hamiltonian hybrid theory}

There is no unique generally accepted theory of interaction between micro and macro degrees
of freedom, where the former are described by quantum and the latter by classical theory.
The reason is primarily because each of the suggested theories has some unexpected or
controversial features (see \cite{Elze} for an informative review). Partial selection of
hybrid theories can be found in \cite{Sudarshan,Boucher,Andronov,Terno,Royal,Hall1}.
Some of the suggested hybrid theories are mathematically inconsistent, and ``no go" type
theorems have been formulated \cite{Salcedo}, suggesting that no consistent hybrid theory
can be formulated. Nevertheless, mathematically consistent but inequivalent hybrid theories
exist \cite{Elze,Hall1,Royal}. The Hamiltonian hybrid theory, as formulated and discussed
for example in \cite{Elze,usPRA}, has many of the properties commonly expected of a good
hybrid theory. In fact, the dynamical formulas of the Hamiltonian theory are equivalent to
the well known mean field approximation, the main novelty being that the theory is formulated
entirely in the framework of the theory of Hamiltonian dynamical systems, which enables
useful insights and methods of analysis \cite{Elze2,usPRA2,FF}. Analysis of cloning in the
Hamiltonian hybrid system is one such application. In fact, we shall analyze possibility
of cloning in general HHM where the Hamiltonian is not necessarily of the mean-field form,
and contrast the results with the HHM of the restricted type where the Hamiltonian is of
the mean-field form.

The phase space in the Hamiltonian theory of a hybrid classical-quantum system, denoted by
${\cal M}$, is considered as a Cartesian product ${\cal M}={\cal M}_c\times {\cal M}_q$ of
the classical subsystem phase space ${\cal M}_c$ with ${\rm dim}\,{\cal M}_c=2N_c$ and of the
quantum subsystem phase space ${\cal M}_q$ with ${\rm dim}\,{\cal M}_q=2N_q$. Local coordinates
on the product are denoted $(q,p,x,y)$, where $(q,p)\in{\cal M}_c$ are called the classical 
degrees of freedom (CDF) and $(x,y)\in{\cal M}_q$ are called the quantum degrees of freedom (QDF).
Notice that the classical and the quantum parts are composed as if both were classical, i.e.,
there is no possibility of entanglement between CDF and QDF. Generalized Hamiltonian hybrid theory
is given by Hamiltonian dynamical system on the phase space ${\cal M}={\cal M}_c\times {\cal M}_q$.
In the general case, nothing is supposed about the total Hamiltonian, and it is only the structure
of the phase space that justifies the terminology of hybrid quantum-classical systems. The Poisson
bracket on $\cal M$ of arbitrary functions of the local coordinates $(q,p,x,y)$ is defined as\vspace{-2mm}
\begin{align}\label{Poisson_hybrid}
\{f_1,f_2\}_{\cal M}&=\sum_{i=1}^{N_{c}}\left(\frac{\partial f_1}{\partial q_i}
\frac{\partial f_2}{\partial p_i}-\frac{\partial f_2}{\partial q_i}
\frac{\partial f_1}{\partial p_i}\right )\nonumber\\
&+\frac{1}{\hbar}\sum_{j=1}^{N_q}\left(\frac{\partial f_1}{\partial x_j}
\frac{\partial f_2}{\partial y_j}-\frac{\partial f_2}{\partial x_j}
\frac{\partial f_1}{ \partial y_j}\right)\!.
\end{align}\vspace{-3mm}\\
Thus, the Hamiltonian form of the hybrid dynamics on $\cal M$ as the phase space reads
\begin{align}\label{eq_hybrid}
&\dot q=\{q,H\}_{\cal M},\quad \dot p=\{p,H\}_{\cal M},\nonumber\\
&\dot x=\{x,H\}_{\cal M},\quad \dot y =\{y,H\}_{\cal M},
\end{align}
where $H$ is an arbitrary smooth function on the total phase space ${\cal M}$.

A particular case of HHM, treated for example in \cite{Elze,usPRA} and equivalent to the mean
field approach, is obtained by further assumptions about the form of the Hamiltonian. The evolution
equations of the hybrid system are in this type of HHM given by the Hamiltonian of the following form
\begin{align}\label{Ham_hybrid}
&H_t(q,p,x,y)=\langle\psi_{x,y}|\hat H_q+\hat V_{int}(q,p)|\psi_{x,y}\rangle+H_c(q,p)\nonumber\\
&=H_c(q,p)+H_q(x,y)+V_{int}(q,p,x,y),
\end{align}
where $H_c$ is the Hamilton's function of the classical subsystem, $H_q(x,y)=\langle\psi_{x,y}|\hat H_q|
\psi_{x,y}\rangle$ is the Hamilton's function of the quantum subsystem and $V_{int}(q,p,x,y)=\langle
\psi_{x,y}|\hat V_{int}(q,p)|\psi_{x,y}\rangle$, where $\hat V_{int}(q,p)$ is a Hermitian operator in
the Hilbert space of the quantum subsystem which depends on the classical coordinates $(q,p)$ and
describes the interaction between the subsystems. Despite the fact that the Hamiltonian is a
quadratic function of QDF (and arbitrary function of CDF) the evolution of the QDF is nonlinear
because of the coupling between QDF and CDF.

It is important to mention the evolution of statistical ensembles of hybrid systems in this type of HHM.
Such an ensemble is described by a probability distribution $\rho(q,p,x,y)$, which evolves by the
Liouville equation with the Hamiltonian (\ref{Ham_hybrid}). The following expression
\begin{align}\label{distr}
&\hat\rho(t)=\int_{\cal M}\rho(q,p,x,y;t)\hat\Pi(x,y) dxdydqdp\nonumber\\
&=\int_{\cal M}\rho_q(x,y;t)\hat\Pi(x,y) dxdy=\int_{\cal M}\hat\rho_{cl}(q,p;t)dqdp,
\end{align}
where $\hat \Pi(x,y)$ is a normalized projector onto the vector $|\psi_{x,y}\rangle$, is a well
defined density matrix, representing a state of the QDF at each $t$. There are many $\rho_q(x,y;t)$
giving the same density matrix $\hat\rho(t)$. From the evolution equation satisfied by (\ref{distr}),
or from (\ref{eq_hybrid}), it is seen that a pure state $|\psi(t)\rangle\langle\psi(t)|$ obtained
from an initial ensemble $\rho(q,p,x,y)=\delta(q-q_0)\delta(p-p_0)\delta(x-x_0)\delta (y-y_0)$ with
CDF (and QDF) in pure states is always a pure state of QDF. The evolution equation satisfied by this
pure state is in the form of (linear) Schr\"odinger equation with the Hamiltonian which is a Hermitian
operator that depends explicitly on $(q(t), p(t))$. On the other hand, if the CDF are initially in a
mixed state, a pure state of the QDF will evolve into a mixed state. Furthermore, it was shown in 
\cite{usPRA2}, that the evolution of a general $\rho(t)$ will involve explicitly the convex 
expansion (\ref{distr}), and not only $\rho(t)$. Therefore, it seems that this type of HHM can be used
for super-luminal communication between distant subparts of the quantum DF.

Discussion of cloning within the restricted type of HHM with the classical part playing the role
of the cloning machine requires special treatment as compared with the general HHM.

\subsection{Definitions of cloning and self-replication}

Cloning is a process involving three systems: the object system $S_o$ with the state space
${\cal S}_o$, a target system $S_t$ with the state space ${\cal S}_t$ the same as that of $S_o$
and an auxiliary system, the cloning machine $S_m$, with the state space of dimension $M$ that
is not specified in advance. It is said that cloning of some arbitrary object state $X_o\in{\cal S}_o$
is possible if there is a state of the target $X_{t,in}\in{\cal S}_{t}$ and a state of the machine
$X_{m,in}\in {\cal S}_m$ such that
\begin{equation}\label{clon}
X_o\odot X_{t,in}\odot X_{m,in}\rightarrow X_o\odot\{X_{t}=X_o\}\odot X_m(X_o).
\end{equation}
The arbitrary state of the object system is conserved by cloning; one fixed 
state of the target and another fixed state of the machine are chosen as 
initial, independently of the object state. The fixed initial target state is 
mapped into the initial state of the object. The final state of the machine 
might depend on the object state $X_o$. It is not assumed that the final machine 
state is uniquely related to $X_o$. Observe that the possibility of cloning does not
imply that the cloning is  achieved with any initial target and machine states, but only
with a specific  choice of these states. The domain and the range of the cloning map
(\ref{clon}) are proper subsets of the sets of possible states of the object+target+machine 
system.

The system $S_o\bigcup S_t\bigcup S_m$ is characterized by its natural evolution,
and the question is if the cloning map belongs to that class. In our case the natural
evolution is given by a Hamiltonian flow on ${\cal S}_o\odot{\cal S}_t\odot{\cal S}_m$,
and thus preserves the symplectic structure on ${\cal S}_o\odot {\cal S}_t\odot {\cal S}_m$.
In NHQM all three systems are quantum and, as was stated in the previous subsection, $\odot$
is the tensor product. In HHM we shall consider the case when the object system and the
target are quantum and the machine is classical. Thus, in this case, $\odot$ between the
machine and object+target is the Cartesian product. Alternatively, which we shall not do,
one could analyze cloning with all three systems of the hybrid nature. The only fixed property
of the cloning problems within the Hamiltonian framework is the canonical Hamiltonian evolution
and the fact that pure states are represented by points in the corresponding phase spaces.
If $X_o$ and $X_t$ represent phase space points in the Hamiltonian formulation corresponding
to the vectors $|\psi_o\rangle$ and $|\psi_t\rangle$, respectively, then it is natural to assume
that the  cloning is successful if at the output $|\psi_t\rangle\exp(i\theta)=|\psi_o\rangle$, i.e.,
\begin{eqnarray}\label{phase}
&& x_t^i\cos\theta-y_t^i\sin\theta=x_o^i,\quad y_t^i\cos\theta+x_t^i\sin\theta=y_o^i,\nonumber\\
&&i=1,2,\dots,N.
\end{eqnarray}

The role of the machine DF can be justified from two different points of view. 
One is the operational point of view, where the appearance of the cloning 
machine is natural. The other role of the cloning machine is to actually enable 
the object+target subsystem to evolve in non-Hamiltonian way. Quite analogously 
to the role of the machine in the standard QM formulation of cloning, here the 
presence of the cloning machine enables the total object+target+machine system 
to evolve canonically while enabling more general type of evolution of the 
subsystem object+target.
In this respect a related more restrictive problem with no cloning machine is 
sometimes considered. Such a process has been termed self-replication, and 
consists in mapping a fixed state of the target system into an arbitrary state 
of the object system, the latter remaining unchanged, but without any influence 
of the third system. In the self-replication process the object+target system is 
considered as isolated. Together with the problem of proper cloning within NHQM 
(with the cloning machine) we shall also analyze the possibility of 
self-replication in such theories.

\section{Main results}

Our strategy to analyze possibility of self-replication and cloning will be the same
in NHQM and HHM. Let us denote by ${\cal M}_{otm}$ the total phase space of the 
object+target+machine system. By ${\cal M}_{in}\subset {\cal M}_{otm}$ we
denote the submanifold of the total phase space of the form ${\cal M}_o\odot 
X_{t,in}\odot X_{m,in}$ where $X_{t,in}$ and $X_{m,in}$ are specific initial 
vectors representing states of the target and the machine, respectively. We shall 
all the time deal with vectors of unit norm. Similarly, we denote by ${\cal M}_{f}
\subset{\cal M}_{otm}$ the submanifold which is the image of ${\cal M}_{in}$ by the
cloning map. Points in ${\cal M}_{f}$ are of the form $X_o\odot X_o\odot X_{m,f}(X_o)$,
$X_o\in {\cal M}_o$, and thus ${\rm dim}\,{\cal M}_{f}={\rm dim}\,{\cal M}_{in}={\rm dim}
\,{\cal M}_o$. We then choose an arbitrary point $X\in {\cal M}_{in}$ and two arbitrary 
normalized tangent vectors $g_X,h_X\in T_X({\cal M}_{in})\subset T_X({\cal M}_{otm})$.
Value of the symplectic area $\omega_X(g_X,h_X)$ is then computed. Cloning
(or self-replication) is represented by the mapping $\phi: {\cal M}_{in}\rightarrow
{\cal M}_{f}$ with the tangent map $\phi_{\star}: T_X({\cal M}_{in})\rightarrow
T_{\phi (X)}({\cal M}_{f})$. Symplectic area between the images of the two vectors
$\omega_{\phi(X)}(\phi_{\star} g_X,\phi_{\star} h_X)$ is then computed. If $\phi$ is
a symplectic map, i.e., can be generated by a piecewise smooth Hamiltonian flow, then
 \begin{equation}\label{symp_cond}
 \omega_{\phi(X)}(\phi_{\star} g_X,\phi_{\star} h_X)= \omega_X(g_X,h_X).
\end{equation}
If (\ref{symp_cond}) is not satisfied, for any choice of $X_{t,in},X_{m,in}$ and 
$X_{m,f}$ then the cloning (self-replication) map $\phi$ cannot be realized by a 
Hamiltonian flow. To apply the procedure, we shall write explicitly the cloning 
map $\phi$ and its tangent map ${\phi}_{\star}$, corresponding to the phase 
spaces ${\cal M}_{in}$ and ${\cal M}_{f}$ with a specific choice of the initial 
target and machine states in the NHQM and HHM. The only difference will be in 
the way the machine phase space ${\cal M}_m$ is added to the phase space of the 
object+target.


In our discussion, we shall consider the simplest possible systems as object, 
target and machine. The object and the target are each taken to be a single qubit.
An arbitrary state of the object qubit is a normalized ${\mathbb C}^2$ vector with
complex coefficients $(\alpha,\beta)$ corresponding to some basis of ${\cal S}_o$.
Furthermore, the initial state of the target qubit will be represented by vector $(1,0)$
in a basis of ${\cal S}_t$ chosen in the same way as the basis in ${\cal S}_o$. This
does not seem to be a restriction with crucial consequences, but grossly simplifies
explicit formulas for the self-replication (and later cloning) map.

In the case of NHQM the machine is also a quantum system and the coupling of it 
with the object+target is via tensor product. In order to demonstrate that in NHQM
cloning by a symplectic (nonlinear) transformation is possible, it is enough to
assume that the cloning machine is also a qubit, set initially in the state
$(\alpha_{m},\beta_m)=(1,0)$, represented in some basis of ${\cal S}_m$. Cloning 
is also possible by a symplectic map in the case of general HHM, when the 
machine is a classical system with two degrees of freedom and is coupled to the
object+target via the Cartesian product. However, an additional argument is used
to show that in the specific HHM with the Hamiltonian of the form (\ref{Ham_hybrid}),
i.e., quadratic in the QDF, cloning of the quantum state is impossible by
symplectic transformation generated by the Hamilton functions of the stated form.

{\it Impossibility of self-replication in NHQM}

Let us first illustrate the computations for the case of self-replication in 
NHQM. Real dimension of ${\cal M}_{in}$ with normalized object states is three.
In the complex notation the initial point in ${\cal M}_{in}$ representing the 
state of object+target before self-replication is
\begin{equation}\label{m_init}
X_{in}=(\alpha,0,\beta,0), \> |\alpha|^2+|\beta|^2=1.
\end{equation}
Two normalized tangent vectors $g$ and $h$ in $T({\cal M}_{in})$ at $X_{in}$ are given as
\begin{subequations}\label{g_init}
\begin{align}
g_{re}&=(-g_1\alpha_{im}+g_3\beta_{re},0,-g_3\alpha_{re}-g_2\beta_{im},0),\\
g_{im}&=(g_1\alpha_{re}+g_3\beta_{im},0,-g_3\alpha_{im}+g_2\beta_{re},0),
\end{align}
\end{subequations}
with arbitrary real numbers $g_1$, $g_2$ and $g_3$ chosen to respect the unity norm. Analogue formulas
apply to $h_{re}$ and $h_{im}$. Subscripts $re$ and $im$ stand for real and
imaginary parts. The skew product of the two tangent vectors is
\begin{equation}\label{symp_init}
\omega(g,h)=(g_3(h_1-h_2)+(g_2-g_1)h_3)(\alpha_{re}\beta_{re}+\alpha_{im}\beta_{im}).
\end{equation}
In formulas (\ref{g_init}) and (\ref{symp_init}) we have, for the sake of brevity,
skipped the subscript indicating the related point $X_{in}$.

Image by the self-replication map $\phi$ of $X_{in}$, again in the complex coordinates, is given by
\begin{equation}\label{m_fin}
X_{f}=(\alpha^2,\alpha\beta,\beta\alpha,\beta^2)\exp(i\theta(\alpha,\beta)).
\end{equation}
Notice the arbitrary phase factor added to the result of the self-replication operation.
Images of $g$ and $h$ by the tangent map $\phi_{\star}$ are given by rather long formulas
which we do not reproduce here. However, the skew product of $\phi_{\star}g$ and $\phi_{\star}h$
at the point $X_f$ is given by
\begin{eqnarray}
\omega(\phi_{\star} g,\phi_{\star} h)&=&2 (g_3(h_1-h_2)+(g_2-g_1)h_3)\times\nonumber\\
&&(\alpha_{re}\beta_{re}+\alpha_{im}\beta_{im})(|\alpha|^2+|\beta|^2).
\end{eqnarray}
Notice that the previous result is independent of arbitrary phase factor. Ratio of
the symplectic areas after and before the application of the self-replication map is
\begin{equation}\label{ratio_1}
\frac{\omega(\phi_{\star} g,\phi_{\star} h)}{\omega(g,h)}=2(|\alpha|^2+|\beta|^2)=2.
\end{equation}
Thus, self-replication map does not preserve the skew product, and therefore cannot be realized
by any symplectic map between ${\cal M}_{in}$ and ${\cal M}_f$.

{\it Possibility of cloning in NHQM}

Consider now the proper cloning map in NHQM with the quantum machine included. 
Since we shall see that the cloning map is symplectic with the cloning machine 
given by a qubit, it is enough to assume this simplest realization of the 
machine. The final state of the machine $(\alpha_{mf},\beta_{mf})$ is free to 
chose, and the choice can be done such that the factor of $2$ appearing in the 
result of self-replication (\ref{ratio_1}) can be canceled.

Formulas for the initial point and its image by the cloning map for the indicated
choice of initial states of the target and the machine, in the complex notation are given by:
\begin{equation}\label{m_init_2}
X_{in}=(\alpha,0,0,0,\beta,0,0,0)
\end{equation}
\begin{eqnarray}\label{m_fin_2}
X_{f}&=&(\alpha^2\alpha_{mf},\alpha^2\beta_{mf},\alpha\beta\alpha_{mf},\alpha\beta\beta_{mf},\nonumber\\
&&\alpha\beta\alpha_{mf},\alpha\beta\beta_{mf},\beta^2\alpha_{mf},\beta^2\beta_{mf}),
\end{eqnarray}
where $(\alpha_{mf},\beta_{mf})$ denote the final state of the machine.
Tangent vector $g$ is given by
\begin{subequations}
\begin{align}
&g_{re}=(-g_1\alpha_{im}+g_3\beta_{re},0,0,0,-g_3\alpha_{re}-g_2\beta_{im},0,0,0),\\
&g_{im}=(g_1\alpha_{re}+g_3\beta_{im},0,0,0,-g_3\alpha_{im}+g_2\beta_{re},0,0,0),
\end{align}
\end{subequations}
and analogously for $h$. The skew product between $g$ and $h$ is
\begin{equation}
\omega(g,h)=(g_3(h_1-h_2)+h_3(g_2-g_1))(\alpha_{re}\beta_{re}+\alpha_{im}\beta_{im}).
\end{equation}
The images of $g$ and $h$ by the tangent map, their skew product, and the ratio
${\omega(\phi_{\star}g,\phi_{\star}h)}/{\omega(g,h)}$ are given by rather long formulas,
which depend on the final machine state. However, we have found that the choice of final
machine state as $(\alpha_{mf},\beta_{mf})=(\bar\alpha,\bar\beta)$, where the bar indicates
complex conjugation, renders the ratio equal to unity for normalized state $(\alpha,\beta)$
of the object. Therefore, the cloning map can be realized by a symplectic transformation.
From the standard QM it follows that the symplectic cloning transformation in NHQM must be nonlinear.

{\it Possibility of cloning in general HHM}

We chose the object and the target to be the same systems and to be in the same 
states as in the case of NHQM. The machine is chosen to be a convenient 
classical system with two DF and coordinates $(q_{1m},q_{2m},p_{1m},p_{2m})$ or 
in complex notation $(q_{1m}+ip_{1m},q_{2m}+ip_{2m})=(\alpha_{m},\beta_{m})$. 
Formulas for the initial point for the indicated special choice of initial 
target and machine states, are given in the complex coordinates by:
\begin{equation}\label{m_init_3}
X_{in}=(\alpha,0,\beta,0,1,0).
\end{equation}
The machine final state is free to choose. With the choice $(\alpha_{mf}=\alpha_{im}+i\alpha_{re},
\beta_{mf}=\beta_{im}+i\beta_{re})$ the state after cloning operation is
\begin{equation}\label{m_fin_3}
X_{f}=(\alpha^2,\alpha\beta,\beta\alpha,\beta^2,\alpha_{im}+i\alpha_{re},\beta_{im}+i\beta_{re}).
\end{equation}
Tangent normalized vector $g$ is given by
\begin{subequations}
\begin{align}
g_{re}&=(-g_1\alpha_{im}+g_3\beta_{re},0,-g_3\alpha_{re}-g_2\beta_{im},0,0,0),\\
g_{im}&=(g_1\alpha_{re}+g_3\beta_{im},0,-g_3\alpha_{im}+g_2\beta_{re},0,0,0),
\end{align}
\end{subequations}
and similarly for tangent vector $h$. Skew product between $g$ and $h$ is given by
\begin{equation}
\omega(g,h)=(g_3(h_1-h_2)+h_3(g_2-g_1))(\alpha_{re}\beta_{re}+\alpha_{im}\beta_{im}).
\end{equation}
The images of the normalized tangent vectors and their skew product are again 
given by rather long formulas. However, the above choice of the machine final 
state renders the ratio
\begin{equation}\label{ratio_3}
\frac{\omega(\phi_{\star}g,\phi_{\star}h)}{\omega(g,h)}=1,
\end{equation}
for normalized initial object states.
Again, the cloning map can be realized by a symplectic transformation.

{\it Impossibility of cloning in the HHM with the specific form of the Hamiltonian}

Special form of the hybrid Hamiltonian (\ref{Ham_hybrid}) implies special status 
of the cloning operation in this type of HHM, as compared with the general case.
In fact, due to the properties of the evolution of pure hybrid states, 
summarized in Section II, pure states of QDF remain pure if the initial state of 
CDF is also pure. Furthermore, the scalar product between two QDF pure states is 
preserved. Therefore, the standard no-cloning argument from linear QM applies. 
Thus, cloning of quantum states is impossible within the specific HHM with 
Hamiltonian (\ref{Ham_hybrid}), and with classical DF assuming the role of the
cloning machine. Here we have an example  of a theory that does not admit cloning
of pure quantum states, but whose natural extension that includes ensembles admits
super-luminal communication.

\section{Discussion}

{\bf Remark 1} {\it Physical interpretation and consequences}: Cloning is 
commonly considered as an information processing task. From this point of view, 
the problem formulated in Section II and discussed in Section III is rather 
formal, and is concerned with idealized system that could never occur in 
information processing protocols with real systems. Pure states of isolated 
systems and their idealized evolution are only probabilistically related to 
information and its processing.
Therefore, relation between the system's states and information must be 
probabilistic, and the processing of such information necessary involves 
stochastic perturbations. This has been analyzed in the standard QM 
\cite{broadcasting}. The question of cloning in real, experimentally available 
systems was not studied in the present publication, but is important in 
analyzing the fundamental and practical consequences.
In order to do that, one needs to use probability ensembles, represented by 
distributions on the relevant phase spaces and stochastic evolution equations. 
We believe that only with such an analysis one could attempt to draw conclusions 
as to the physical consistency of the nonlinear HQM and HHM.

{\bf Remark 2} {\it Cloning vs.\ super-luminal signaling}: It is well known that 
if the cloning would be possible in the standard QM then, also in the framework 
of this theory, it would also be possible to communicate information at 
super-luminal speed. It has also been claimed that the condition of no 
super-luminal signaling puts an upper bound on the fidelity of cloning, in 
effect excluding the perfect cloning in QM \cite{Gisin2}.
The condition of no super-luminal signaling is in \cite{Gisin2} expressed in 
terms of convex expansions of mixed states. In the opposite direction, it has 
been argued \cite{Gisin, Mielnik} that a nonlinear evolution of pure quantum 
states would enable signaling at super-luminal speed. This is consistent with
our results which show the possibility of cloning in NHQM. 
However, the argument does not exclude theories in which pure quantum states 
cannot be perfectly cloned, but the super-luminal signaling is possible. 
Mean-field HHM with the special form of the Hamiltonian (\ref{Ham_hybrid}) is an 
example of such a theory.


{\bf Remark 3} {\it Cloning in classical mechanics}: It is commonly understood 
that perfect cloning of classical information contained in a classical pure 
state is possible. Of course, in order to discuss the possibility of cloning,
one needs precise definition of the state space and the type of dynamics
characterizing the classical system. One formulation of the problem, 
particularly relevant in fundamental physics and for comparison with our 
results, is for the classical system modeled using the framework of classical 
Hamiltonian dynamical systems. States of the system, the target and the machine 
are described by the corresponding symplectic manifolds, their union is given by 
the Cartesian product and the symplectic structure on the total space is such 
that the symplectic structures on the components are obtained by the 
corresponding projections. It is known that the self-replication is not, but the 
cloning is possible by symplectic mappings on the total phase space, provided 
that the machine space has enough dimensions \cite{classical}. The proof of no 
self-replication is similar to the case in nonlinear quantum mechanics, 
presented in Section III.
Possibility of cloning in Hamiltonian CM is established and discussed by concrete examples 
of symplectic cloning maps. It should be stressed that cloning is performed by 
linear symplectic mapping. On the other hand, cloning in NHQM and general HHM 
can be achieved by a symplectic map which must be nonlinear. This seems to 
be the crucial difference between the theories involving tensor or Cartesian 
products between the target and the object systems.

{\bf Remark 4} {\it Cloning in classical statistical mechanics}: Evolution of a 
probability distribution generated by a measure preserving mapping of a phase 
space is by definition linear, and preserves the relative entropy between two 
distributions. This two properties, i.e., preservation of a nontrivial (quasi) 
distance between states and linearity are features of the Schr\"odinger 
evolution of pure quantum states. Also, the space of statistical states of a 
compound system, for example $L^1({\cal M}_1\times {\cal M}_2)$ can be 
considered as the tensor product of $L^1({\cal M}_1)$ and $L^1({\cal M}_2)$. 
Thus, all three ingredients that are used in the standard proofs of no-cloning 
in QM are also properties of classical statistical mechanics. Therefore, one 
expects, and it has been proved to be true \cite{stat}, that cloning in 
classical statistical mechanics is impossible. Due to the creation of 
correlations between the subsystems, it is also possible to formulate the 
question of cloning in a more general way, more akin to the notion of broadcasting 
in QM. The answer to the question of possibility of broadcasting in Hamiltonian
CM is also negative \cite{stat}.\\

\section{Summary}

We have analyzed possibility of exact cloning of unknown quantum states in two 
types of nonlinear generalizations of quantum mechanics. Both types of 
generalizations were formulated as Hamiltonian dynamical systems on appropriate 
phase spaces. In the first type, which we called nonlinear Hamiltonian quantum 
mechanics (NHQM), the object, the target and the machine are treated as quantum 
systems, and it is shown that cloning can be realized by a nonlinear symplectic 
mapping. On the other hand, the process of self-replication, involving only the 
system and the target, cannot be realized by any symplectic transformation in 
NHQM. The other type of nonlinear generalizations of QM, which we have treated 
describes hybrid quantum-classical systems, again using the framework of 
Hamiltonian dynamical systems. Here, the object and the target are quantum, but 
the machine is a classical system. We have shown that there exists a nonlinear 
symplectic transformation which realizes the cloning operation. However, the 
cloning transformation cannot be realized in the Hamiltonian hybrid theory of 
the mean-field type, in which case the Hamiltonian must be a quadratic function 
of the quantum degrees of freedom and an arbitrary one of the classical degrees 
of freedom. It would be interesting to try to extend these results onto the 
problem of broadcasting of mixed states in the nonlinear generalizations. This 
would require analysis of the Liouville evolution of densities and might result 
in possibility of broadcasting also in the mean-field Hamiltonian hybrid theory.

\begin{acknowledgments}
We acknowledge support of the Ministry of education, science and technological 
development of the Republic of Serbia, under contracts No.\ 171006, 171017,
171020, 171038 and 45016 and COST (Action MP1006).
\end{acknowledgments}

\end{document}